\documentclass[referee,pdflatex,sn-nature]{sn-jnl}
\usepackage{amsmath,amssymb,amsfonts}
\usepackage{graphicx}
\usepackage{amsthm}
\usepackage{verbatim}
\usepackage{epigraph}

\def\be{\begin{equation}}
\def\ee{\end{equation}}
\def\ber{\begin{eqnarray}}
\def\eer{\end{eqnarray}}
\def\bern{\begin{eqnarray*}}
\def\eern{\end{eqnarray*}}

\def\dv{\mathbf{d}}

\def\jv{\mathbf{j}}

\def\qv{\mathbf{q}} 

\def\Rv{\mathbf{R}} 

\def\Vv{\mathbf{V}}

\def\0v{\mathbf{0}}
\def\1v{\mathbf{1}}
\def\2v{\mathbf{2}}
\def\3v{\mathbf{3}}

\def\jv{\mathbf{j}}

\def\pa{\partial}

\DeclareMathAlphabet\mathbfcal{OMS}{cmsy}{b}{n}

\def\Im{ {\rm Im} \, }

\begin{document}

\title{Viscous AC current-driven nanomotors}

\author*[1]{\fnm{Vladimir U.} \sur{Nazarov}}\email{vladimir.nazarov@mail.huji.ac.il}

\author[2]{\fnm{Tchavdar N.} \sur{Todorov}}\email{t.todorov@qub.ac.uk}

\author[1]{\fnm{E.~K.~U.} \sur{Gross}}\email{eberhard.gross@mail.huji.ac.il}

\affil[1]{\orgdiv{Fritz Haber Research Center of Molecular Dynamics, Institute of Chemistry}, \orgname{the Hebrew University of Jerusalem}, \city{Jerusalem},  \country{Israel}}

\affil[2]{\orgdiv{School of Mathematics and Physics}, \orgname{Queen's University Belfast}, \city{Belfast},  \country{UK}}

\abstract{
{\large The recent discovery that electrons in nano-scale conductors can act like a highly viscous liquid has triggered a surge of research activities investigating consequences of this surprising fact. Here we demonstrate that the electronic viscosity has an enormous influence on the operation of a prototypical AC-current-driven nano-motor. The design of this prototype consists of a diatomic molecule immersed in an otherwise homogeneous electron liquid which carries an AC current. The motion of the diatomic is determined by a subtle balance between the current-induced forces and electronic friction. By ab-initio time-dependent density-functional simulations we demonstrate that the diatomic performs a continuous rotation provided the amplitude and frequency of the imposed AC current lie within certain islands of stability. Outside these islands the nuclear motion is either chaotic or comes to a stand-still. The proposed design of the nano-motor is the conceptually simplest realization of the idea of an molecular waterwheel sandwiched between conducting leads.}}

\maketitle

\newpage

\begin{flushright}
\noindent{\it And yet it moves}

\noindent Galileo Galilei
\end{flushright}

The idea of current-driven nano-motors took off in the 00’s with proposals for molecular-scale windmills, waterwheels and related concepts 
\cite{Seideman-03,Kral-05,Bailey-08,Seldenthuis-10}. These devices hinge on the transfer of angular momentum from the electric current to atoms or groups of atoms as a result of chirality \cite{Kral-05,Bailey-08}  or, more generally, of the non-conservative nature of current-induced forces \cite{Dundas-09,Lu-10,Mishima-25}.

Recently the molecular electronics community has been jolted by the realization that electrons in a conductor can behave as a highly viscous fluid \cite{Polini-20}.
Electron viscosity, as a manifestation of the dynamic many-body dissipative effects in three- and low-dimensional conductors \cite{Giuliani&Vignale}, has an immediate bearing on such phenomena as the electrical resistivity \cite{Sai-05,Koentopp-06,Nazarov-14-2}, current-induced forces on nuclei \cite{Nazarov-24}, and slowing down of ions in matter \cite{Nazarov-05,Nazarov-07}.
These processes are all-important and closely interdependent in the functionality of current-driven nano-motors.

Here we demonstrate the operation of an AC molecular motor in viscous electron liquid. 
We show for the first time that electron viscosity has a quantitative and qualitative significance for these devices, to the extent that it can make the difference between the molecular motor working or not working.

\section*{Theoretical framework}

We consider two nuclei, of the charges $Z_1$, $Z_2$ and masses  $M_1$,$M_2$,  immersed in the otherwise homogeneous electron gas (HEG) of density $\bar{n}$,
and subject to the action of a uniform AC current with density $\overline{\jv} (t)$.
Adopting the picture of weak nuclei-HEG interaction, denoting by $\Rv_c(t)$ and $\Rv_r(t)$ the instantaneous position of the center of mass (c.m.) and the relative position of the two nuclei, respectively, 
in the {\it Methods} section we show  that  the motion of the nuclei are governed by the system of coupled non-linear differential equations
\begin{equation}
\hspace{-1.1 cm}
\begin{split}
  \ddot{\Rv}_c(t)
& \! = \!    \frac{  (Z_1 \! + \! Z_2) }{M_c \bar{n}}  \dot{\overline{\jv}} 
-   \frac{2 }{\pi M_c} \int  \frac{Q(q) \qv}{q^4} \left[Z_1^2+Z_2^2  +2 Z_1 Z_2  e^{-i \qv\cdot \Rv_r(t) } \right]  
\left\{\left[\frac{\overline{ \jv}(t)}{\bar{n}}- \dot{\Rv}_c(t) \right]\cdot \qv\right\} d\qv   \\
& \! + \!  \frac{2 }{\pi M_c^2 } \! \int  \! \frac{Q(q) \qv}{q^4} 
\left[  M_1 Z_2^2 \! - \! M_2 Z_1^2   \! + \! Z_1 Z_2 (M_1 \! - \! M_2) e^{-i \qv\cdot \Rv_r(t) } \right] \left[  \dot{\Rv}_r(t) \! \cdot \! \qv \right] \! d\qv  \! - \! \frac{2 K_c}{M_c} \Rv_c(t),  
\end{split}
\label{vcr14}
\end{equation}
\begin{equation}
\hspace{-1.1 cm}
\begin{split}
  \ddot{\Rv}_r(t)
& \! = \!    \frac{1 }{  \bar{n}}  \left( \frac{Z_2}{M_2} \! - \! \frac{Z_1}{M_1} \right) \dot{\overline{ \jv}} (t) 
\! +  \! \frac{2 }{\pi M_c } \! \int \!  \frac{Q(q) \qv}{q^4} \left[   \dot{\Rv}_r(t) \! \cdot \! \qv \right]
\left[ \frac{M_1 Z_2^2}{M_2 } \! + \! \frac{M_2 Z_1^2}{M_1 }   \! - \! 2 Z_1 Z_2 e^{i \qv\cdot \Rv_r(t) } \right] \!  d\qv \\
&-   \frac{2 }{\pi} \int  \frac{Q(q) \qv}{q^4}  \left\{\left[\frac{\overline{\jv}(t)}{\bar{n}}- \dot{\Rv}_c(t) \right]\cdot \qv\right\}
\left[\frac{Z_2^2}{M_2} -\frac{Z_1^2}{M_1} + \left(\frac{1}{M_2}-\frac{1}{M_1} \right)  Z_1 Z_2 e^{i \qv\cdot \Rv_r(t) } \right] d\qv \\
&+     \frac{2}{ i \pi} \left(\frac{1}{M_1}+\frac{1}{M_2} \right) \int   \frac{\qv}{q^4} Z_1 Z_2 \chi^h(q,0) e^{i \qv\cdot \Rv_r(t)} d\qv 
+   \left(\frac{1}{M_1}+ \frac{1}{M_2}\right)  \frac{Z_1  Z_2 \Rv_r(t)}{|\Rv_r(t)|^3}.
\end{split}
\label{vcr24}
\end{equation}
Here $M_c=M_1+M_2$, $\chi^h(q,\omega)$ is the wave-vector and frequency-dependent density response function of the HEG \cite{Giuliani&Vignale},  $Q(q)$ is its  wave-vector-resolved friction coefficient, defined by Eq.~\eqref{intw0} of the  {\it Methods}, and the overhead dot stands for time differentiation.
The physical interpretation of the terms in Eqs.~\eqref{vcr14}-\eqref{vcr24} is the following:
the first terms on the RHS of both equations, those involving $\dot{\overline{ \jv}} (t)$, are accelerations due to the direct force from the field on the nuclei;
the terms involving $\overline{\jv}(t)$ are  due to the current-induced forces; the terms involving the velocities are the friction decelerations;
the last two terms in Eq.~\eqref{vcr24} are due to the  restoring force  toward the equilibrium separation between the nuclei; 
and the last term in Eq.~\eqref{vcr14} represents a harmonic force introduced
to confine the c.m. of the impurity to the origin, where $K_c$ is its
stiffness. In a real-world molecular junction where the molecule is
sandwiched between the atomic tips of metallic leads, the confinement of
the c.m. is achieved by chemical bonds to the tip atoms, sufficiently
diffuse to allow for the rotation of the molecule, but strong enough
to confine the molecule between the tips of the leads. 

By numerically solving the system \eqref{vcr14}-\eqref{vcr24} we simulate the motion of the diatomic impurity in HEG and, in particular, identify the conditions for the operation of the waterwheel.

\section*{Results}

We have conducted calculations for the impurity comprised of  a proton and a deuteron immersed in HEG of $r_s=2$, $6$, and $10$ a.u, where $r_s$ is the density parameter defined as $1/\bar{n}= (4/3) \pi r_s^3$.

\begin{figure}[h!]
\includegraphics[width=\columnwidth, clip=true, trim=61 5 19 8]{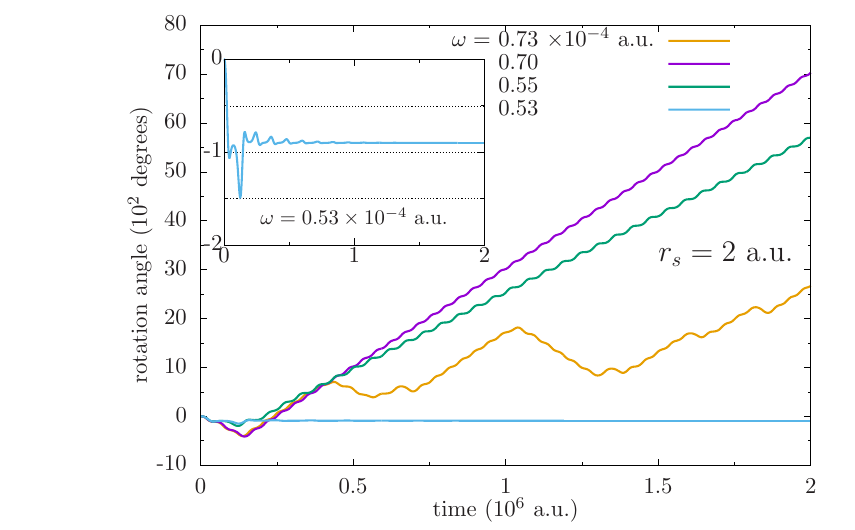}
\caption{\label{rs2wdep}
Angle between the instantaneous direction of the axis of the impurity  $\Rv_r(t)$ and its initial value $\Rv_r(0)$ versus time.
Starting from $t=0$, the current density $\bar{\jv}(t)=\bar{\jv}_0 \sin \omega t$ is applied with $\bar{\jv}_0$ perpendicular to $\Rv_r(0)$ and $|\bar{\jv}_0|= 4 \times 10^{-5}$ a.u.
At selected current frequencies, exemplified by $\omega=0.70 \times 10^{-4}$ and $0.55 \times 10^{-4}$ a.u., a continuous rotation of the impurity (the wavy straight  lines in the graph) takes place, while it is suppressed for $\omega= 0.73\times 10^{-4}$ and $0.53\times 10^{-4}$ a.u.
Inset shows separately, on a magnified scale, the same time evolution for  $\omega=0.53 \times 10^{-4}$ a.u., where rotation stalled at the angle of $-90^\circ$ is observed. 
}
\end{figure}

In Fig.~\ref{rs2wdep}, the angle between the instantaneous direction of the axis of the impurity  and its initial direction along the $x$-axis is plotted versus the elapsed time,
for a fixed value of the current-density amplitude and for four values of the frequency.
We observe that whether the continuous rotation  takes place (slightly wavy straight lines in the graph) or not depends on the value of the frequency $\omega$ of the applied current at the given amplitude $|\bar{\jv}_0|$.
At the allowed {\em resonant} values of the frequency, the rotational motion stabilizes at approximately constant angular velocity equal to the frequency  of the applied current (the cases of $\omega=0.70 \times 10^{-4}$ and $0.55 \times 10^{-4}$ a.u. in Fig.~\ref{rs2wdep}). By contrast, at  the off-resonance frequencies, the motion is either chaotic or comes to a halt, as is the case in Fig.~\ref{rs2wdep} for $\omega=0.73 \times 10^{-4}$ and $0.53 \times 10^{-4}$ a.u., respectively.
The resonant bands have been studied for the closely related mathematical and engineering problem of the damped driven non-linear pendulum \cite{Clifford-95}, to which we will return later.
The supporting material provides a video contrasting the motion of the impurity within and outside a resonance band.

Importantly, the continuous rotation never occurs with precisely a constant angular velocity: it can be verified analytically that Eqs.~\eqref{vcr14}-\eqref{vcr24} do not admit a monochromatic solution. 
The same can be seen in  Fig.~\ref{rs2wdep},
where results for the resonant frequencies $\omega=0.70 \times 10^{-4}$ and $0.55 \times 10^{-4}$ a.u.
exhibit a weakly oscillatory character.
Quantitative analysis reveals that the frequency of the superimposed oscillations is twice the frequency $\omega$ of the applied current.
The origin of these oscillations can be understood as  follows: During one revolution of the wheel, there are two maximal pushes on it, when the molecular axis passes  the  direction perpendicular to the current, and there are two `dead zones' with minimal driving, when the molecular axis and the current are parallel. This accounts for the double-$\omega$ oscillations in the rotation speed.

\begin{figure}[h!]
\includegraphics[width=\columnwidth, clip=true, trim=47 4 15 7]{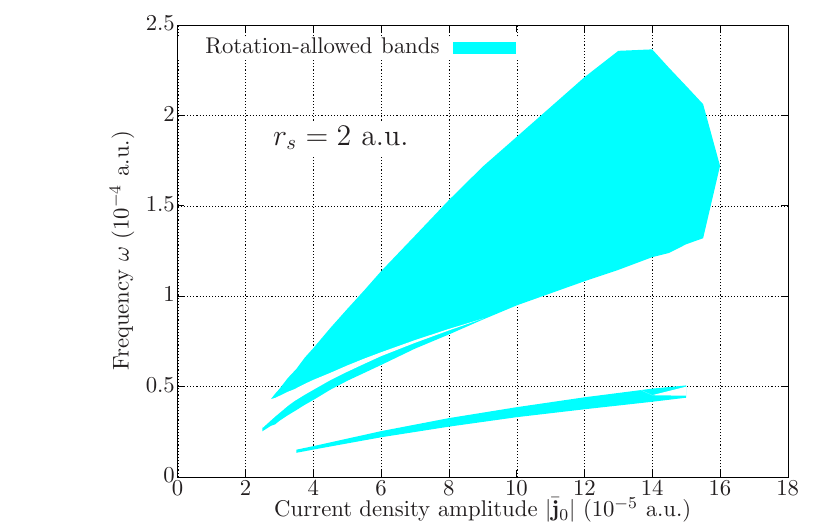}
\caption{\label{ggrs2}
Phase-diagram, in the current density amplitude -- frequency coordinates, for HEG of $r_s=2$ a.u. Within the painted areas (bands), a continuous rotation persists, while it is forbidden outside of it.
}
\end{figure}

\begin{figure}[h!]
\includegraphics[width=\columnwidth, clip=true, trim=53 0 15 0]{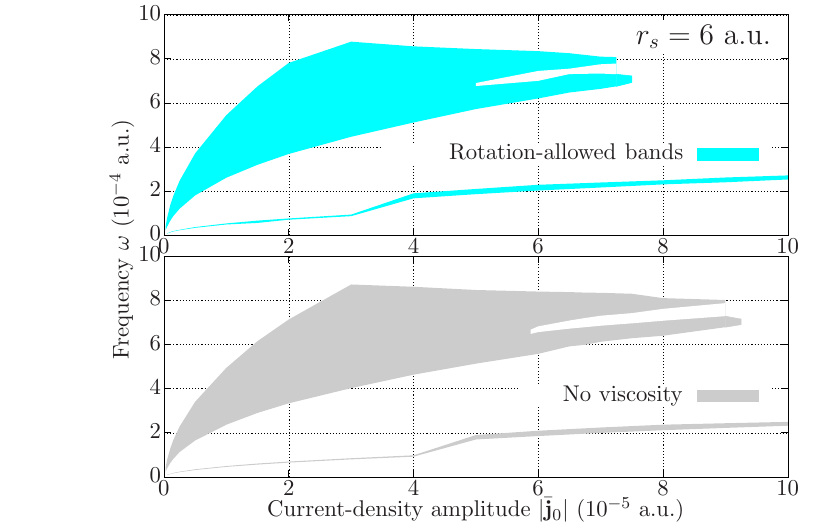}
\caption{\label{ggrs6}
Same as Fig.~\ref{ggrs2}, but $r_s=6$ a.u. 
The lower panel shows the phase diagram  with  neglect of the viscosity contribution [$\Im f_{xc}(q,\omega)$ set to zero].
}
\end{figure}

\begin{figure}[h!]
\includegraphics[width=\columnwidth, clip=true, trim=47 4 15 0]{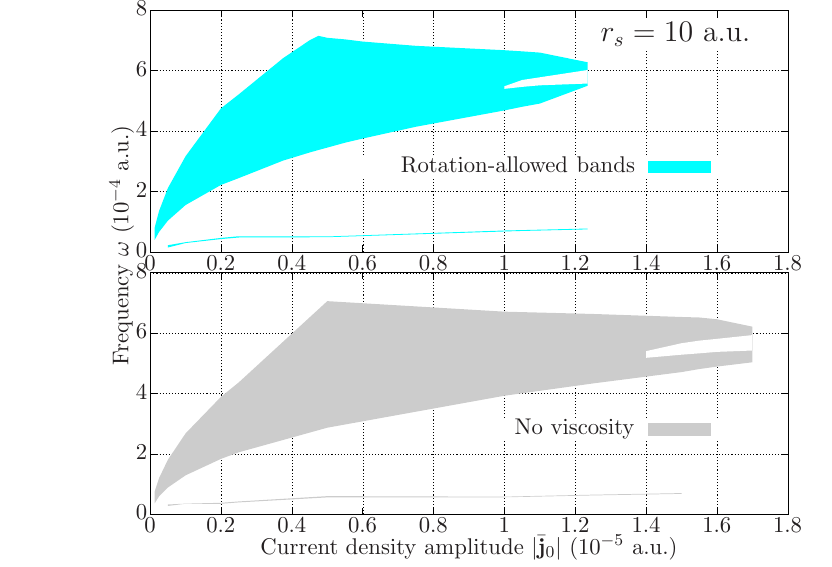}
\caption{\label{ggrs10}
Same as Fig.~\ref{ggrs6}, but $r_s=10$ a.u.
}
\end{figure}

In order to quantify the ranges of the rotation-allowed amplitudes and frequencies, in Figs.~\ref{ggrs2}, \ref{ggrs6}, and \ref{ggrs10}
we present the $|\bar{\jv}_0|-\omega$ phase-diagrams for HEG of $r_s=2$, $6$, and $10$ a.u., respectively.
The painted areas correspond to the pairs  of the current-density amplitude $|\bar{\jv}_0|$ and the frequency $\omega$ which support the rotation ({\it resonant bands}).
Outside those areas, the rotation is suppressed. 
The allowed bands are dictated by a balance between the current-induced forces and the electronic friction.
It is within those amplitude-frequency regions that the waterwheel is functional.
The position of the bands and their configuration depend crucially  on the value of the $r_s$-parameter,
indicating the critical  dependence of the motion of the impurity on the density of HEG.  

For more dilute HEG ($r_s=6$ and $10$ a.u. in the figures), the {\em electron viscosity}, which is a dynamic many-body property manifested via the imaginary part of the exchange-correlation kernel $f^h_{xc}(q,\omega)$ \cite{Giuliani&Vignale}, starts to play a role. In Figs.~\ref{ggrs6} and \ref{ggrs10}, we compare the rotation-allowed bands calculated with and without  account of the viscosity [$\Im f^h_{xc}(q,\omega)$ is set to zero in the latter case]. The importance of the viscous contribution to the forces is particularly clear in the case of $r_s=10$ a.u. (Fig.~\ref{ggrs10}), where the neglect of the viscosity leads to a considerable overestimation of the area of the allowed band, together with a larger spread towards  the higher-current domain.

\section*{Rotating pendulum model}

It is instructive to consider a physically transparent model, which can give a good approximation to the full theory based on Eqs.~\eqref{vcr14}-\eqref{vcr24}. This can be done assuming a rigid bond of the molecule during rotation, i.e., assuming $|\Rv_r(t)|=d=const$. As shown in the {\it Methods}, in this case the angle of the rotation $\theta(t)$ obeys the equation of motion of the rotating pendulum
\begin{equation}
\begin{split}
\ddot{\theta}(t)+b \dot{\theta}(t) &= \left[A \sin \omega t +B \omega \cos \omega t \right] 
  [j_{0x} \sin \theta(t)-j_{0y} \cos \theta(t)].
\end{split}
\label{pend1}
\end{equation}
Here $b$ is the friction coefficient, and $A$ and $B$ are two driving force factors, i.e., the current-induced one and the direct force, respectively.
Their explicit expressions in terms of the HEG quantities are given by Eqs.~\eqref{pend21}-\eqref{pend4} of the {\it Methods}.

\begin{figure}[h!]
\includegraphics[width=\columnwidth, clip=true, trim=63 0 14 0]{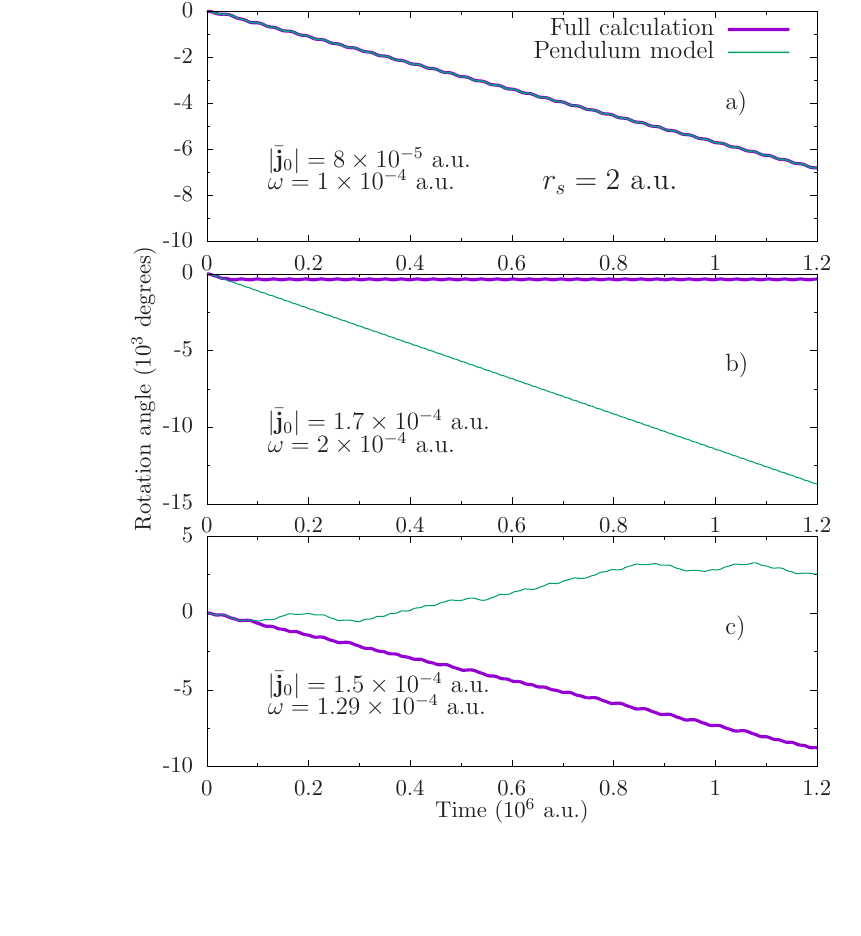}
\caption{\label{pend}
Angle of rotation of the molecule versus time. Results of the full calculations by Eqs.~\eqref{vcr14}-\eqref{vcr24} are compared with those of the pendulum model of Eqs.~\eqref{pend1}-\eqref{pend4}. 
}
\end{figure}

\begin{figure}[h!]
\includegraphics[width=\columnwidth, clip=true, trim=26 0 0 0]{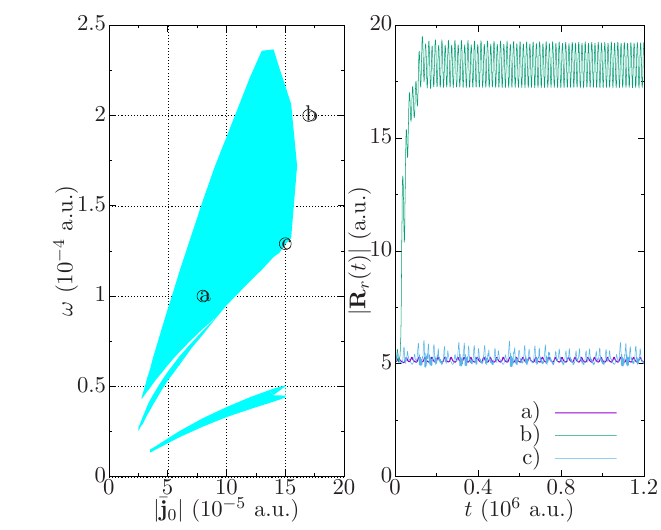}
\caption{\label{ggrs2dots}
Left: Positions of the $(|\bar{\bf{j}}_0|,\omega)$ points, corresponding to the results of the calculations in Fig.~\ref{pend},  on the phase diagram of Fig.~\ref{ggrs2}. Right: Corresponding time-dependence of the distance between the two nuclei.
}
\end{figure}

In Fig.~\ref{pend}, results of the full calculations [those by solving Eqs.~\eqref{vcr14}-\eqref{vcr24}] are compared with the solution of Eq.~\eqref{pend1} for three points in the current-density amplitude - frequency plane. Those points are marked in the left panel of Fig.~\ref{ggrs2dots},
and the corresponding time dependence of the bond lengths are shown in the right panel of the same figure.
In case a), which is well inside the rotation-allowed band, the full and the pendulum-model rotations are indistinguishable from each other. This is in accord with the separation between nuclei remaining almost constant and equal to its equilibrium value (right panel of Fig.~\ref{ggrs2dots}).

Case b) exemplifies the dissociation of the molecule, which occurs in the full calculation, with a cessation of the rotation. The pendulum model, however, predicts a continuous rotation. The variance here is not surprising since the pendulum model does not include  dissociation. 

Interestingly, case c) demonstrates the reverse situation, where the  radial motion (`breathing' of the molecule) stabilizes the continuous rotation in the full calculation, while the pendulum model predicts erroneously a chaotic motion.

\section*{Conclusion and outlook}

 We have developed a theory of the motion of a diatomic impurity
in an electron liquid, wherein the dynamics of electrons is handled within the linear response theory, while  the motion of nuclei is treated non-perturbatively.
This method provides a means to describe the continuous rotation of the impurity under the action of an AC current, which process is shown to be impossible  within  the pure linear response  theory. By the use of the new theory, a diatomic waterwheel propelled by an AC current has been conceptually and computationally constructed.

The simplicity of the HEG model notwithstanding, a wealth of physical phenomena reveals itself in the motion of a compound impurity immersed in this medium. Firstly, we have found that whether the waterwheel is functional or not depends on definite conditions. Namely, to make it work, the amplitude and the frequency of the applied current must fall into the {\em bands} of allowed magnitudes, the rotation being prohibited otherwise. Those bands are formed due to the intricate balance between the accelerating current-induced forces and the decelerating electronic friction.

Secondly, by applying an advanced modern theory of excitations in the electron liquid, we have accurately accounted for both the single-particle and multi-electron  effects.
The latter are known to involve the electron {\em viscosity}, which, as we show, plays an important part in the waterwheel operation, affecting  the rotation-allowed bands both quantitatively and qualitatively.

Demonstrating and overcoming  the fundamental inability of the purely linear theory to  describe the operation of a turbine in electron liquid, 
our results advance the theoretical foundations of the field of nano-motors in a new direction. 

\bmhead{Acknowledgements}
This project has received funding from the European Research Council (ERC) under the European Union's Horizon 2020 research and information programme (Grant Agreement No. ERC-2017-AdG-788890). 


\section*{Methods}

\subsection*{Equations of motion}

The problem of the motion of a diatomic impurity in the homogeneous electron gas (HEG) under the action of current-induced forces (electron wind) and electronic friction has been earlier addressed and solved  within the consistent {\em linear response} approach, where `consistent' here indicates that both the excitation of the HEG and the displacements of nuclei from their  equilibrium positions were treated to the first order in the externally applied current \cite{Nazarov-24}. The linear response theory is well suited for the description of the vibrational and translational  motion  under the action of a weak perturbation. Indeed, the former is due to small amplitudes of vibrations under  weak fields. For the latter, although the displacement of the center of mass (c.m.) of a molecule may grow large with time, its coordinates do not enter equations of motion, but only its velocity does, which is a consequence of the translational invariance of the HEG system \cite{Nazarov-24}.

\begin{figure}[h!]
\includegraphics[width=0.85 \columnwidth, clip=true, trim=70 30 70 45]{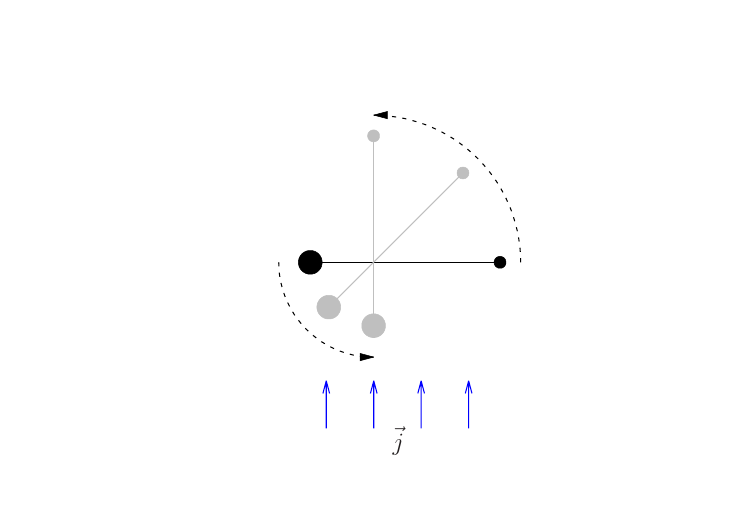}
\caption{\label{rot1}
Schematics of a diatomic impurity rotation under  current-induced force. 
Whatever weak the current is, the atoms' displacements cannot remain small in the continuous rotation regime.
}
\end{figure}

The situation  differs cardinally in the case of the rotational motion. Even under a weak current, the continuous rotation consists in  the repeated full revolutions of the wheel. 
The latter suggests the continuous change of the orientation of the axis of the  molecule relative to the current direction, which change cannot be considered small (see a schematic illustration in Fig.~\ref{rot1}). As a consequence, only the initial stage of the rotation can be caught within the linear response theory.
The subject of this paper being  the waterwheel, we have, necessarily, to go beyond the linear response regime with respect to the  displacements of nuclei.

Our starting point is the equations of motion of the diatomic impurity in HEG, as they read in the linear response regime \cite{Nazarov-24} [atomic units ($\hbar=e^2=m_e=1$) are used  throughout unless otherwise indicated]
\begin{equation}
\hspace{-3 cm}
\begin{split}
\omega     \Vv_c(\omega)
& \! = \!   \frac{ \omega (Z_1 \! + \! Z_2) }{M_c \bar{n}}  \overline{  \jv} (\omega)  
\! - \!  \frac{2}{\pi M_c\omega} \! \int  \! \frac{\qv}{q^4}  \left[ \chi^h(q,\omega) \! - \! \chi^h(q,0)\right] \left[Z_1^2 \! + \! Z_2^2  \! + \! 2 Z_1 Z_2  e^{-i \qv\cdot \dv } \right]  
\left\{\left[\frac{\overline{  \jv}(\omega)}{\bar{n}} \! -  \! \Vv _c(\omega)\right] \! \cdot \! \qv\right\} \! d\qv   \\
&+   \frac{2}{\pi M_c^2 \omega} \! \int  \! \frac{\qv}{q^4}  \left[ \chi^h(q,\omega)-\chi^h(q,0)\right] 
\left[  M_1 Z_2^2-M_2 Z_1^2   +Z_1 Z_2 (M_1-M_2) e^{-i \qv\cdot \dv } \right] [ \Vv_r(\omega)\cdot \qv] d\qv ,  
\end{split}
\label{vcr1}
\end{equation}
\begin{equation}
\hspace{-3 cm}
\begin{split}
\omega    \Vv_r(\omega)
& \! = \!  \frac{   \omega}{  \bar{n}} \!  \left( \frac{Z_2}{M_2} \! - \! \frac{Z_1}{M_1} \right) \overline{\jv} (\omega)  
  \! + \! \! \frac{2}{\pi M_c \omega} \! \int \!  \frac{\qv}{q^4}  \left[ \chi^h(q,\omega) \! - \! \chi^h(q,0)\right] [    \Vv _r(\omega) \! \cdot \! \qv]
\left[ \frac{M_1 Z_2^2}{M_2 }\!  + \! \frac{M_2 Z_1^2}{M_1 }   \! -\!  2 Z_1 Z_2 e^{i \qv\cdot \dv } \right]  \! d\qv \\
&-   \frac{2}{\pi \omega} \int  \frac{\qv}{q^4}  \left[ \chi^h(q,\omega)-\chi^h(q,0)\right] \left\{\left[\frac{\overline{  \jv}(\omega)}{\bar{n}}-  \Vv _c(\omega)\right]\cdot \qv\right\}
\left[\frac{Z_2^2}{M_2} -\frac{Z_1^2}{M_1} + \left(\frac{1}{M_2}-\frac{1}{M_1} \right)  Z_1 Z_2 e^{i \qv\cdot \dv } \right] d\qv \\
&-    \frac{2}{\pi \omega} \left(\frac{1}{M_1}+\frac{1}{M_2} \right) \int   \frac{\qv}{q^4} Z_1 Z_2 \chi^h(q,0) e^{i \qv\cdot \dv }
 [\Vv_r(\omega)\cdot \qv] d\qv  
- \frac{1}{ \omega} \left(\frac{1}{M_1}+ \frac{1}{M_2}\right) [   \Vv_r(\omega)\cdot \nabla_{\dv}] \nabla_{\dv} \frac{Z_1  Z_2}{d} ,
\end{split}
\label{vcr2}
\end{equation}
were $\omega$ is the frequency of the externally applied  monochromatic current-density $\overline{\jv} (\omega)$, $\Vv_c(\omega)$ and $\Vv_r(\omega)$ are the velocities of the c.m. of the molecule and that of the relative motion of its constituent nuclei, respectively,
$Z_\alpha$ and $M_\alpha$, $\alpha=1,2$, are the charges and masses of the nuclei, respectively, $M_c=M_1+M_2$, 
$\bar{n}$ is the density of HEG, $\dv$ is the equilibrium relative position of the nuclei at rest, and $\chi^h(q,\omega)$ is the wave-vector and frequency-dependent density response function of the HEG \cite{Gross-85,Giuliani&Vignale}.

A major progress towards the construction of the nonlinear theory of the nuclear motion can be achieved by noting that, for nuclei, heavy as they are in comparison with electrons,  only the low-frequency part of the electronic excitation spectrum plays a significant role. 
Indeed, we have seen that our frequencies of interest are of the order of $10^{-4}$ a.u., which is small compared 
to the characteristic plasma frequency $\omega_p$ of the considered HEG (e.g., $\omega_p=0.61$, $0.12$, and $0.055$ a.u., at $r_s=2$, $6$, and $10$ a.u., respectively).
This justifies  a  substitution to be made  in Eqs.~\eqref{vcr1}-\eqref{vcr2}
\begin{equation}
\begin{split}
\frac{1}{i} \frac{\chi^h(q,\omega)-\chi^h(q,0)}{\omega}  \to
 \left. \frac{\pa \Im  \chi^h(q,\omega)}{\pa \omega}\right|_{\omega=0}  =  Q(q),
\end{split}
\label{intw0}
\end{equation}
where $Q(q)$, defined by Eq.~\eqref{intw0}, can be viewed as the wave-vector resolved friction coefficient of the HEG.
After the substitution~\eqref{intw0} (and only with it), Eqs.~\eqref{vcr1}-\eqref{vcr2} can be readily Fourier-transformed to the time domain, which leads to Eqs.~\eqref{vcr14}-\eqref{vcr24}.
We note that {(I) For an additional generality, in Eq.~\eqref{vcr14} we have confined the c.m. by the harmonic restoring  force $-2 K_c \Rv_c(t)$, where, as a specific case, $K_c$ can be zero; (II) Although, for brevity, in Eqs~\eqref{vcr14}-\eqref{vcr24} we have kept imaginary exponents, it can be shown that RHSs of these equations are purely real.} 

We emphasize, and this is key, that, parallel to  transferring to the time-domain, in Eqs~\eqref{vcr14}-\eqref{vcr24} we have substituted the {\em equilibrium} relative position of the nuclei $\dv$ with the {\it instantaneous} one $\Rv_r(t)=\Rv_2(t)-\Rv_1(t)$, which is justifiable, again, owing to the different time-scales  of the nuclear and electronic motions. The latter substitution has made Eqs~\eqref{vcr14}-\eqref{vcr24}  non-linear with respect to the motion of nuclei, while the electron dynamics is still treated within the linear response theory. 

\subsection*{Calculational procedures and the response functions used}
The system of coupled ODE  \eqref{vcr14}-\eqref{vcr24} is to be solved to determine the trajectories $\Rv_c(t)$ and $\Rv_r(t)$ (we could, of course,  return to the individual coordinates  $\Rv_1(t)$ and $\Rv_2(t)$, if desired).
The coupling between Eqs.~\eqref{vcr14} and \eqref{vcr24} reflects the fact that the c.m. motion of the molecule does not separate from the relative one,
which is due to the mediation by HEG and is in contrast to the situation in vaccum \cite{Nazarov-24}.

The  molecule being at its  equilibrium  at $t\le 0$, at $t>0$ we subject it to the uniform  monochromatic current-density
\begin{equation}
\bar{\jv}(t)=\bar{\jv}_0 \sin \omega t,
\label{j0}
\end{equation}
and perform the time-propagation. The density response function $\chi^h(q,\omega)$ of HEG, entering equations \eqref{vcr14}-\eqref{vcr24}, is obtained from the relation \cite{Gross-85}
\begin{equation}
1/\chi^h(q,\omega)=1/\chi^h_s(q,\omega)-4 \pi/q^2 -f_{xc}^h(q,\omega),
\end{equation}
where $\chi^h_s(q,\omega)$ is the Lindhard's independent-electron density response function \cite{Lindhard-54} and $f_{xc}^h(q,\omega)$ is the exchange-correlation (xc) kernel of  HEG \cite{Gross-85}. For the latter, we use the constraint-based approximation rMCP07, which is currently considered accurate at all densities of the fluid phase of HEG \cite{Kaplan-22}.

Equations~\eqref{vcr14}-\eqref{vcr24} were solved using the variable-step Runge-Kutta integrator after Tsitouras and Papakostas \cite{Tsitouras-99,williams_rklib}. In all the calculations, we have been setting $K_c=0.55$ a.u, which ensures c.m. of the molecule to be pinned at the origin by the harmonic force  in Eq.~\eqref{vcr14}.

We clarify that, while the calculation for a given pair $(\bar{\jv}_0,\omega)$ is deterministic, producing of the phase-diagrams of Figs.~\ref{ggrs2}-\ref{ggrs10} is partly heuristic. Indeed, to determine the band edges, the divide and conquer algorithm was employed. This involved, at a given $\bar{\jv}_0$, scanning over a grid of $\omega$-s. Then, for any two adjacent values of the latter found, belonging to the allowed and forbidden bands, the band edge point was determined by consecutive divisions of the intervals by halves. Obviously, some extra very narrow bands might have been overlooked by this procedure. We, however, believe that all the main bands are presented in the figures.

\subsection*{Particulars of the rotating pendulum model}

Solutions of Eqs~\eqref{vcr14}-\eqref{vcr24} involve both the tangential and radial motion of the nuclei. It is, however, instructive to isolate the cases when the bond $|\Rv_r(t)|$ remains approximately fixed, which leads to the rotating pendulum model for the impurity's motion. We will be seeking for the approximate solution of the form
\begin{align}
&\Rv_c(t)=\0v, \label{Rvc0}\\
&\Rv_r(t)=d \left[ \cos \theta(t),\sin \theta(t) \right],
\end{align}
where $d$ is the equilibrium distance between the nuclei. Then
\begin{equation}
\dot{\Rv}_r(t) = d \, \dot{\theta}(t) \left[- \sin \theta(t),\cos \theta(t) \right]
\end{equation}
and
\begin{equation}
\begin{split}
\ddot{\Rv}_r(t) =d \, \ddot{\theta}(t) \left[- \sin \theta(t),\cos \theta(t) \right] 
- d [\dot{\theta}(t)]^2 \left[ \cos \theta(t),\sin \theta(t) \right].
\end{split}
\label{Rvpp0}
\end{equation}
Substituting Eqs.~\eqref{Rvc0}-\eqref{Rvpp0} into Eq.~\eqref{vcr24} and separating the tangential component (the one parallel to $\left[- \sin \theta(t),\cos \theta(t) \right]$), we arrive at  Eq.~\eqref{pend1} for the rotation angle $\theta(t)$, where
\begin{equation}
\begin{split}
b =  \frac{8  }{ M_c }  \int\limits_0^\infty d q Q(q) 
\left[ \frac{2 Z_1 Z_2 }{d^3 q^3} (\sin q d - q d \cos qd) 
 - \frac{1}{3} \left(\frac{M_1 Z_2^2}{M_2 } + \frac{M_2 Z_1^2}{M_1 }\right)    \right],
\end{split}
\label{pend21}
\end{equation}
\begin{equation}
\begin{split}
A =\frac{8 }{ \bar{n} d} \int\limits_0^\infty  d q   Q(q) 
\left[\frac{1}{3}  \left(\frac{Z_2^2}{M_2} -\frac{Z_1^2}{M_1}\right) 
 + \frac{Z_1 Z_2 }{d^3 q^3} \left(\frac{1}{M_2}-\frac{1}{M_1} \right)  (\sin q d - q d \cos qd) \right],
\end{split}
\end{equation}
\begin{equation}
B= \frac{1 }{ \bar{n} d}  \left( \frac{Z_1}{M_1}-\frac{Z_2}{M_2} \right).
\label{pend4}
\end{equation}
We emphasize that, in the derivation of  Eq.~\eqref{pend1}, we have ignored the radial component of Eq.~\eqref{vcr24}, 
the equation for which is incompatible with the assumption of the rigid bond. We have, however, seen that the latter effect may occur weak, justifying 
the introduction of the simple pendulum model.

\end{document}